\documentclass[12pt]{article}%
\usepackage{amsmath}
\usepackage{amsfonts}
\usepackage{amssymb}
\usepackage{graphicx}%
\begin{document}
\leftline {USC-06/HEP-B2 \hfill hep-th/0605267}{\small \
}{\vskip-1cm}

.

{\vskip2cm}

\begin{center}
{\Large \textbf{Interacting Two-Time Physics Field Theory}}

{\Large \textbf{With a BRST Gauge Invariant Action}}\footnote{This work was
partially supported by the US Department of Energy under grant number
DE-FG03-84ER40168.}{\Large \textbf{\ }}

{\vskip0.8cm}

\textbf{Itzhak Bars and } \textbf{Yueh-Cheng Kuo}

{\vskip0.8cm}

\textsl{Department of Physics and Astronomy}

\textsl{University of Southern California,\ Los Angeles, CA 90089-2535 USA}

{\vskip1.5cm} \textbf{Abstract}
\end{center}

We construct a field theoretic version of 2T-physics including interactions in
an action formalism. The approach is a BRST formulation based on the
underlying Sp$\left(  2,R\right)  $ gauge symmetry, and shares some
similarities with the approach used to construct string field theory. In our
first case of spinless particles, the interaction is uniquely determined by
the BRST gauge symmetry, and it is different than the Chern-Simons type theory
used in open string field theory. After constructing a BRST gauge invariant
action for 2T-physics field theory with interactions in $d+2$ dimensions, we
study its relation to standard 1T-physics field theory in $\left(  d-1\right)
+1$ dimensions by choosing gauges. In one gauge we show that we obtain the
Klein-Gordon field theory in $\left(  d-1\right)  +1$ dimensions with unique
SO$\left(  d,2\right)  $ conformal invariant self interactions at the
classical field level. This SO$\left(  d,2\right)  $ is the natural linear
Lorentz symmetry of the 2T field theory in $d+2$ dimensions. As indicated in
Fig.1, in other gauges we expect to derive a variety of SO$\left(  d,2\right)
$ invariant 1T-physics field theories as gauge fixed forms of the same 2T
field theory, thus obtaining a unification of 1T-dynamics in a field theoretic
setting, including interactions. The BRST gauge transformation should play the
role of duality transformations among the 1T-physics holographic images of the
same parent 2T field theory. The availability of a field theory action opens
the way for studying 2T-physics with interactions at the quantum level through
the path integral approach.

{\newpage}

\section{Sp$\left(  2,R\right)  $ gauge symmetry and 2T-physics}

The essential ingredient in 2T-physics is the basic gauge symmetry Sp$(2,R)$
acting on phase space $X^{M},P_{M}$. Under this gauge symmetry, momentum and
position are locally indistinguishable, so the symmetry leads to some deep
consequences. Some of the phenomena that emerge include certain types of
dualities, holography and emergent spacetimes.

The simplest model of 2T-physics is defined by the worldline action
\cite{2treviews}\footnote{The simplest 2T-physics action is in flat $d+2$
spacetime. More generally 2T-physics is defined in the presence of arbitrary
background fields, including electromagnetism, gravity and high spin fields
\cite{2tbacgrounds}.}
\begin{equation}
S_{2T}=\frac{1}{2}\int d\tau~D_{\tau}X_{i}^{M}X_{j}^{N}\eta_{MN}%
\varepsilon^{ij}=\int d\tau~\left(  \dot{X}^{M}P^{N}-\frac{1}{2}A^{ij}%
X_{i}^{M}X_{j}^{N}\right)  \eta_{MN}. \label{2Taction}%
\end{equation}
Here $X_{i}^{M}=\left(  X^{M}\left(  \tau\right)  ,~P^{M}\left(  \tau\right)
\right)  ,$ $i=1,2,$ is a doublet under Sp$\left(  2,R\right)  $ for every
$M,$ the structure $D_{\tau}X_{i}^{M}=\partial_{\tau}X_{i}^{M}-A_{i}^{~j}%
X_{j}^{M}$ is the Sp(2,R) gauge covariant derivative, Sp(2,R) indices are
raised and lowered with the antisymmetric Sp$\left(  2,R\right)  $ metric
$\varepsilon^{ij},$ and the symmetric $A^{ij}\left(  \tau\right)  $ is the
gauge field. In the last expression an irrelevant total derivative $-\left(
1/2\right)  \partial_{\tau}\left(  X\cdot P\right)  $ is dropped from the action.

The Sp$(2,R)$ gauge symmetry renders the solutions for $X^{M}\left(
\tau\right)  ,P^{M}\left(  \tau\right)  $ trivial unless there are two
timelike dimensions. Therefore the target spacetime metric $\eta^{MN}$ must
have $\left(  d,2\right)  $ signature. So, the two timelike dimensions is not
an input, rather it is an output of the gauge symmetry. Sp$\left(  2,R\right)
$ is just sufficient amount of gauge symmetry to remove ghosts due to the two
timelike dimensions, therefore for a ghost free theory with Sp$\left(
2,R\right)  $ gauge symmetry one cannot admit more than two timelike
dimensions. Although the 2T theory is in $d+2$ dimensions, there is enough
gauge symmetry to compensate for the extra $1+1$ dimensions, so that the
physical (gauge invariant) degrees of freedom are equivalent to those
encountered in 1T-physics in $\left(  d-1\right)  +1$ dimensions.

One of the strikingly surprising aspects of 2T-physics is that a given $d+2$
dimensional 2T theory descends, through Sp$\left(  2,R\right)  $ gauge fixing,
down to a family of holographic 1T images in $\left(  d-1\right)  +1$
dimensions, all of which are gauge equivalent to the parent 2T theory and to
each other. However, from the point of view of 1T-physics each image appears
as a different dynamical system with a different Hamiltonian. Fig.1 below
illustrates a family of holographic images that have been obtained from the
simplest model of 2T-physics \cite{2tHandAdS}. The central circle represents
the 2T action in Eq.(\ref{2Taction}), while the surrounding ovals represent
examples of 1T dynamical systems in $\left(  d-1\right)  +1$ dimensions that
emerge from the same theory. The 1T systems include interacting as well as
free systems in 1T-physics.

Hence 2T-physics can be viewed as a unification approach for one-time physics
(1T-physics) systems through higher dimensions. It is distinctly different
than Kaluza-Klein theory because there are no Kaluza-Klein towers of states,
but instead there is a family of 1T systems with duality type relationships
among them.%

%TCIMACRO{\FRAME{dtbpFUX}{5.9819in}{4.4944in}{0pt}{\Qcb{Fig.1 - Some 1T-physics
%systems that emerge from the solutions of Q$_{ij}=0.$}}{\Qlb{Fig1}%
%}{2tphysics061.wmf}{\special{ language "Scientific Word";  type "GRAPHIC";
%width 5.9819in;  height 4.4944in;  depth 0pt;  original-width 5.0004in;
%original-height 3.7498in;  cropleft "0";  croptop "1";  cropright "1";
%cropbottom "0";  filename '2Tphysics061.wmf';file-properties "XNPEUR";}}}%
%BeginExpansion
\begin{center}
\fbox{\includegraphics[
height=4.4944in,
width=5.9819in
]%
{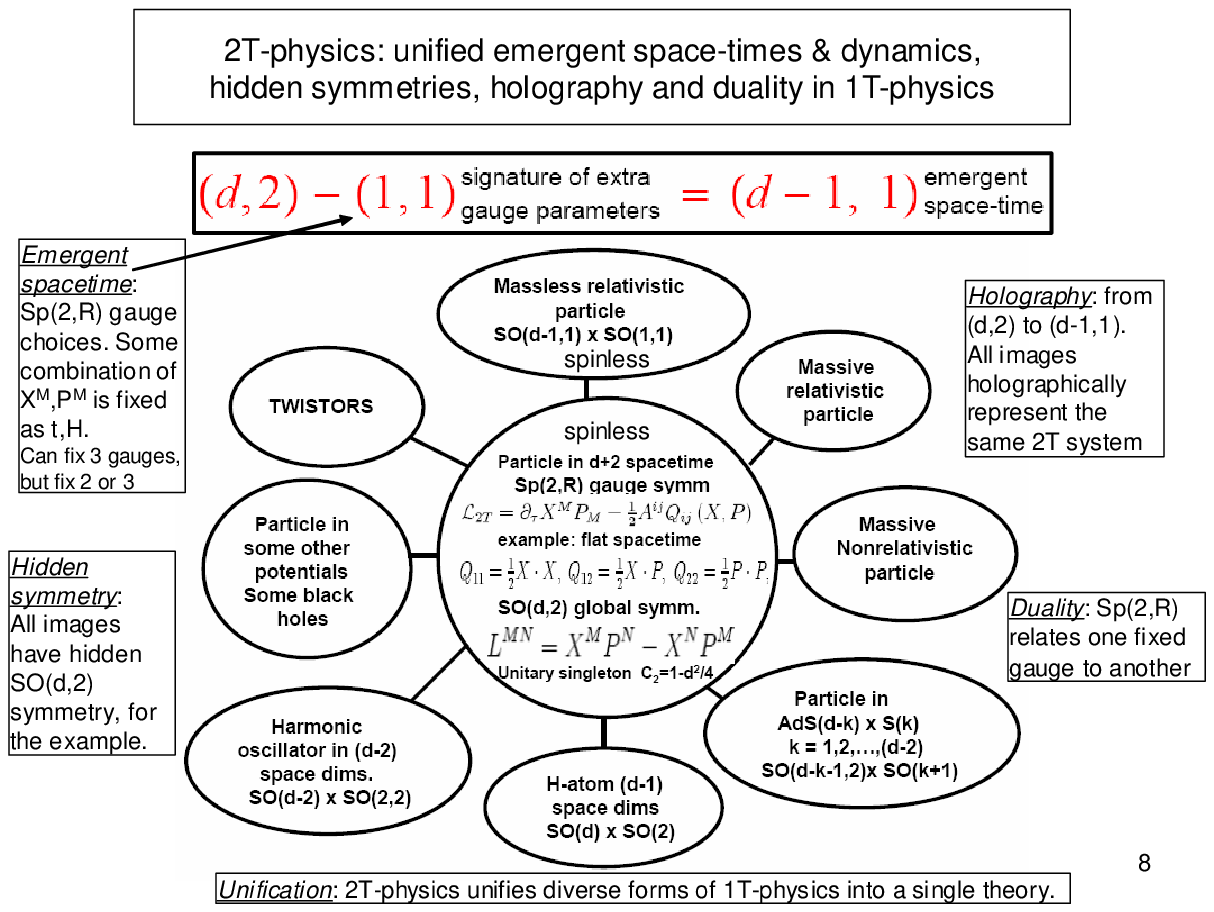}%
}\\
Fig.1 - Some 1T-physics systems that emerge from the solutions of Q$_{ij}=0.$%
\label{Fig1}%
\end{center}
%EndExpansion

In this paper we are interested in constructing a field theoretic formulation
of 2T-physics to explore the field theoretic counterpart of the type of
phenomena summarized in Fig.1. Previous field theoretic efforts in 2T-physics
with interactions are described in \cite{2tfield}\cite{2tfieldXP}. Here we
will use the results of \cite{2tfield} as a guide for self-consistent
equations of motion with interactions, to build an action principle by using a
BRST approach similar to the one used to construct string field theory
\cite{witten}. The BRST action opens the way for discussing the 2T quantum
field theory via the path integral.

\section{Covariant quantization and field equations \label{cov}}

The generators of Sp$\left(  2,R\right)  $ at the quantum level are
$Q_{ij}=\frac{1}{2}X_{(i}\cdot X_{j)}.$ Their explicit form in terms of
canonical variables $X_{i}^{M}=\left(  X^{M}~P^{M}\right)  $ in $\left(
d+2\right)  $ dimensions is
\begin{equation}
Q_{11}=\frac{1}{2}X^{2},\;\;Q_{12}=Q_{21}=\frac{1}{4}\left(  X\cdot P+P\cdot
X\right)  ,\;\;Q_{22}=\frac{1}{2}P^{2}. \label{Qij}%
\end{equation}
They can be rewritten in terms of the standard notation for Sp$\left(
2,R\right)  =SL\left(  2,R\right)  $ generators $J_{m}=\left(  J_{0}%
,J_{1},J_{2}\right)  $ as
\begin{equation}
J_{0}=\frac{1}{4}\left(  P^{2}+X^{2}\right)  ,\;J_{1}=\frac{1}{4}\left(
P^{2}-X^{2}\right)  ,\;J_{2}=\frac{1}{4}\left(  X\cdot P+P\cdot X\right)
\label{Jm}%
\end{equation}
Here $J_{0}$ is the compact generator. The Lie algebra that follows from the
canonical commutation rules $\left[  X^{M},P^{N}\right]  =i\eta^{MN}$ is%
\begin{equation}
\left[  J_{m},J_{n}\right]  =i\varepsilon_{mnk}J^{k},\;\text{metric }\eta
^{mn}=diag\left(  -1,1,1\right)  ,\;\varepsilon_{012}=+1,\;J^{k}=\eta
^{kl}J_{l}.
\end{equation}

The covariant quantization of the theory is defined by the physical states
that are annihilated by the Sp$\left(  2,R\right)  $ generators $J_{m}%
|\phi\rangle=0$. This means that the physical states
%TCIMACRO{\TEXTsymbol{\vert}}%
%BeginExpansion
$\vert$%
%EndExpansion
$\phi\rangle$ are Sp$\left(  2,R\right)  $ gauge invariant. In position space
the probability amplitude $\phi_{0}\left(  X\right)  \equiv\langle
X|\phi\rangle$ must satisfy%
\begin{equation}
\text{ }J_{m}\phi_{0}\left(  X\right)  =0\;\text{or \ }Q_{ij}\phi_{0}\left(
X\right)  =0.
\end{equation}
Taking into account that $P_{M}$ acts like a derivative $P_{M}\phi_{0}\left(
X\right)  =-i\frac{\partial\phi_{0}\left(  X\right)  }{\partial X^{M}}$, we
obtain the following differential equations that determine the physical
states
\begin{equation}
X^{2}\phi_{0}\left(  X\right)  =0,\;\left(  X\cdot\frac{\partial}{\partial
X}+\frac{\partial}{\partial X}\cdot X\right)  \phi_{0}\left(  X\right)
=0,\;\frac{\partial}{\partial X}\cdot\frac{\partial}{\partial X}\phi
_{0}\left(  X\right)  =0. \label{constDiff}%
\end{equation}
The first equation has the solution $\phi_{0}\left(  X\right)  =\delta\left(
X^{2}\right)  A\left(  X\right)  ,$ where $A\left(  X\right)  $ is defined up
to additional terms proportional to $X^{2}$ (since those vanish when
multiplied with $\delta\left(  X^{2}\right)  $). In examining the remaining
equations we take into account the following properties of the delta function
as a distribution (i.e. under integration with smooth functions of $X^{M}$ in
$d+2 $ dimensions)%
\begin{align}
X^{2}\delta\left(  X^{2}\right)   &  =0\label{delta1}\\
X\cdot\frac{\partial}{\partial X}\delta\left(  X^{2}\right)   &  =2X^{2}%
\delta^{\prime}\left(  X^{2}\right)  =-2\delta\left(  X^{2}\right)
\label{delta2}\\
\frac{\partial}{\partial X}\cdot\frac{\partial}{\partial X}\delta\left(
X^{2}\right)   &  =2\left(  d+2\right)  \delta^{\prime}\left(  X^{2}\right)
+4X^{2}\delta^{\prime\prime}\left(  X^{2}\right)  =2\left(  d-2\right)
\delta^{\prime}\left(  X^{2}\right)  \label{delta3}%
\end{align}
Then, in $d+2$ dimensions the remaining differential equations in
Eq.(\ref{constDiff}) reduce to the form%
\begin{equation}
\delta\left(  X^{2}\right)  \left(  X\cdot\frac{\partial A}{\partial X}%
+\frac{d-2}{2}A\right)  =0,\;\;\delta\left(  X^{2}\right)  \frac{\partial
^{2}A}{\partial X_{M}\partial X^{M}}+4\delta^{\prime}\left(  X^{2}\right)
\left(  X\cdot\frac{\partial A}{\partial X}+\frac{d-2}{2}A\right)  =0.
\label{diffdelta}%
\end{equation}
Therefore $A\left(  X\right)  $ must satisfy the following differential
equations%
\begin{equation}
\left(  X\cdot\frac{\partial A}{\partial X}+\frac{d-2}{2}A\right)  _{X^{2}%
=0}=0,\;\;\left(  \frac{\partial^{2}A}{\partial X_{M}\partial X^{M}}\right)
_{X^{2}=0}=0. \label{2Tfield}%
\end{equation}
up to additional terms proportional to $X^{2}$ . The derivatives in these
expressions must be taken before applying $X^{2}=0.$

As indicated above there is freedom in the choice of $A\left(  X\right)  ,$
its first derivative $X\cdot\frac{\partial A}{\partial X}$ and its second
derivative $\frac{\partial^{2}A}{\partial X_{M}\partial X^{M}}$ up to
arbitrary smooth functions proportional to $X^{2}.$ This freedom amounts to a
gauge symmetry. If this freedom is used so that the extra terms are all gauge
fixed to be zero, then the equations above indicate that the gauge fixed
$A\left(  X\right)  $ must be homogeneous of degree $-\frac{d-2}{2}$ and must
satisfy the Klein-Gordon equation in $d+2$ dimensions.

As shown in \cite{2tfield}, there are many ways of parameterizing $X^{M}$ in
$\left(  d,2\right)  $ dimensions that solve $X^{2}=0$. The 1T-physics
physical interpretation of the physical states $\phi\left(  X\right)  $
depends on which of the components of $X^{M}$ is taken to parameterize
\textquotedblleft time\textquotedblright\ in 1T-physics. One of these cases
was discussed by Dirac \cite{Dirac}, and shown to correspond to the massless
particle described by the Klein-Gordon equation in $\left(  d-1,1\right)  $
dimensions. But as shown in \cite{2tfield} there are many other choices of
\textquotedblleft time\textquotedblright\ that yield other differential
equations which describe a variety of dynamical systems in 1T-physics in
$\left(  d-1,1\right)  $ dimensions. Some of the emerging 1T systems are shown
in Fig.1.

As an illustration we show here how the massless Klein-Gordon equation in
$\left(  d-1,1\right)  $ emerges. We use the 2T flat metric $\eta
^{MN}=diag(\overset{0^{\prime}}{-1},\overset{1^{\prime}}{+1},\overset{0}%
{-1},\overset{1}{1}\cdots,\overset{d-1}{1})$ in $\left(  d,2\right)  $
dimensions, and define the lightcone type coordinates $X^{\pm^{\prime}}%
=\frac{1}{\sqrt{2}}\left(  X^{0^{\prime}}\pm X^{1^{\prime}}\right)  $ so that
$X^{M}X^{N}\eta_{MN}=-2X^{+^{\prime}}X^{-^{\prime}}+X^{\mu}X_{\mu}.$ We
parametrize $X^{M}$ in the form
\begin{equation}
X^{+^{\prime}}=\kappa,\;X^{-^{\prime}}=\kappa\lambda,\;X^{\mu}=\kappa x^{\mu}.
\label{massless}%
\end{equation}
The solution to $X^{2}=X^{M}X^{N}\eta_{MN}=0$ is then given by%
\begin{equation}
\lambda=\frac{x^{2}}{2}. \label{massless2}%
\end{equation}
Then we compute the spacetime metric $ds^{2}=dX^{M}dX^{N}\eta_{MN}$ in the
space of solutions of $X^{2}=0,$ and find the emergent spacetime as follows
\begin{align}
ds^{2}  &  =-2dX^{+^{\prime}}dX^{-^{\prime}}+dX^{\mu}dX^{\nu}\eta_{\mu\nu
},\;\eta_{\mu\nu}=diag\left(  -1,1,\cdots,1\right)  ,\\
&  =-2\left(  d\kappa\right)  \left(  \frac{x^{2}}{2}d\kappa+\kappa x^{\nu
}dx_{\nu}\right)  +\left(  x^{\mu}d\kappa+\kappa dx^{\mu}\right)  \left(
x^{\nu}d\kappa+\kappa dx^{\nu}\right)  \eta_{\mu\nu}\\
&  =\kappa^{2}dx^{\mu}dx^{\nu}\eta_{\mu\nu}.
\end{align}
The metric $ds^{2}=dX^{M}dX^{N}\eta_{MN}$ and the constraint $X^{M}X_{M}=0$
are both invariant under SO$\left(  d,2\right)  ,$ so the emergent spacetime
metric $ds^{2}=\kappa^{2}dx^{\mu}dx^{\nu}\eta_{\mu\nu}$ must have the same
symmetry in a hidden way and non-linearly realized on the remaining
coordinates $\kappa,x^{\mu}$. Indeed this is the well known non-linear
SO$\left(  d,2\right)  $ conformal transformations of $x^{\mu}$ under which
$dx^{\mu}dx^{\nu}\eta_{\mu\nu}$ transforms into itself up to an $x$-dependent
conformal factor. Hence $\kappa$ transforms by the inverse factor so that
$ds^{2}=\kappa^{2}dx^{\mu}dx^{\nu}\eta_{\mu\nu}$ remains invariant.

Now we return to the differential equations in Eq.(\ref{2Tfield}) and analyze
them in the parametrization of Eq.(\ref{massless}). The homogeneity condition
is solved in the form
\begin{equation}
A\left(  X^{M}\right)  =\left(  X^{+^{\prime}}\right)  ^{-\left(  d-2\right)
/2}F\left(  \frac{X^{-^{\prime}}}{X^{+^{\prime}}},\frac{X^{\mu}}{X^{+^{\prime
}}}\right)  ,\;\text{assuming }X^{+^{\prime}}\neq0,\label{AtoF}%
\end{equation}
for any function $F.$ Inserting this into the second equation in
(\ref{2Tfield}), first taking the derivatives $\partial/\partial X^{M}$, and
then inserting the form of $X^{M}$ in Eq.(\ref{massless}) gives the
Klein-Gordon equation for the field $\varphi\left(  x\right)  $ in $x^{\mu}$
space in $\left(  d-1\right)  +1$ dimensions\footnote{Here we apply the chain
rule, $\frac{\partial}{\partial X^{M}}=\frac{\partial\kappa}{\partial X^{M}%
}\frac{\partial}{\partial\kappa}+\frac{\partial\lambda}{\partial X^{M}}%
\frac{\partial}{\partial\lambda}+\frac{\partial x^{\nu}}{\partial X^{M}}%
\frac{\partial}{\partial x^{\nu}},$ that follows from Eq.(\ref{massless}) and
then set $\lambda=x^{2}/2$ as in Eq.(\ref{massless2}). Then, $\left(
\frac{\partial^{2}A}{\partial X^{M}\partial X_{M}}\right)  _{X^{2}=0}=\frac
{1}{\kappa^{2}}\frac{\partial^{2}\varphi}{\partial x^{\mu}\partial x_{\mu}}$
where Eq.(\ref{AtoF}) has been used. See section (4.1) in \cite{2tfield} for
details.} (i.e. in the reduced space with one fewer time and one fewer space
dimensions)
\begin{equation}
A\left(  X^{M}\right)  _{X^{2}=0}=\kappa^{-\left(  d-2\right)  /2}%
\varphi\left(  x\right)  ,\;\;\varphi\left(  x\right)  \equiv F\left(
\frac{x^{2}}{2},x\right)  ,\;\partial^{\mu}\partial_{\mu}\varphi\left(
x\right)  =0.\;\;\label{AX}%
\end{equation}
Hence this form of $A\left(  X\right)  ,$ which is fully described by the free
Klein-Gordon field $\varphi\left(  x\right)  $ in $\left(  d-1\right)  +1$
dimensions, is a holographic image of the $d+2$ dimensional 2T-physics system
described by the three equations in (\ref{constDiff}).

We have seen that the $d+2$ dimensional field equations in Eq.(\ref{constDiff}%
) can be recast as field equations in $\left(  d-1\right)  +1$ dimensions.
Depending on the choice of parametrization used to solve the first two
equations, the third equation is cast to a second order equation with a
particular choice of time (embedded in $d+2)$ that controls the dynamical
evolution. This remaining equation corresponds to the wave equation of the
quantum system described by one of the surrounding ovals in Fig.1 (with some
quantum ordering subtleties). For example, instead of the Klein-Gordon
equation, the Laplacian on AdS$_{d-k}\times$S$^{k}$ emerges as discussed in
\cite{2tfield} (this reference contains several more examples). So the set of
differential equations in Eq.(\ref{constDiff}) play the role of the central
circle in Fig.1, and exhibit the type of phenomena summarized in the figure,
namely emergent spacetime, duality, holography, unification, now in a setting
of free field theory rather than classical or quantum mechanics. This result
\cite{2tfield} is at the level of equations of motion, and without interactions.

Our aim in this paper is to propose an action principle that yields the system
of equations in (\ref{constDiff}) as a result of minimizing the action, and
then to generalize the action by including interactions in the 2T field theory
formalism. This will provide an environment to explore the properties of
2T-physics, including emergent spacetime, duality, holography, unification, in
the presence of interactions, and also provide the proper framework for
quantizing the interacting 2T-physics field theory through the path integral approach.

\section{BRST operator for Sp$\left(  2,R\right)  $}

The BRST approach is equivalent to the covariant quantization of the system as
discussed in the previous section, but it can be used to help develop a
natural framework for constructing an action. The following BRST framework is
in position space which is different in detail than the BRST framework in
phase space previously discussed in \cite{2tfieldXP}. These ideas ar following
discussion inspired by the success of the BRST approach in string field theory
\cite{witten}. We will see that the free field part of the action has the same
form as free string field theory, but the interaction will differ from the
Chern-Simons form of string field theory, due to the differences in the
structure of the constraints $J_{m}$ that appear in the BRST operator.

To perform the BRST quantization of the theory we introduce the ghosts
$\left(  c^{m},b_{m}\right)  ,$ with $m=0,1,2$ with the canonical structure%
\begin{equation}
\left\{  c^{m},b_{n}\right\}  =\delta_{~n}^{m},\text{ therefore }b_{n}%
=\frac{\partial}{\partial c^{n}}\text{ on functions of }c^{m}.
\end{equation}
The BRST operator that satisfies $Q^{2}=0$ is
\begin{equation}
Q=c^{m}J_{m}-\frac{i}{2}\varepsilon^{mnk}c_{m}c_{n}b_{k}.\;
\end{equation}

We introduce the field in position space $\Phi\left(  X^{M},c^{m}\right)  $
including ghosts, and expand it in powers of $c^{m}$
\begin{align}
\Phi\left(  X,c\right)   &  =\phi_{0}\left(  X\right)  +c^{m}\chi_{m}\left(
X\right)  +\left(  cc\right)  _{m}\phi^{m}\left(  X\right)  +\left(
ccc\right)  \chi_{0}\left(  X\right) \label{bigF}\\
\left(  cc\right)  _{m}  &  \equiv\frac{1}{2}\varepsilon_{mnk}c^{n}%
c^{k},\;\;\left(  ccc\right)  \equiv\frac{1}{3!}\varepsilon_{mnk}c^{m}%
c^{n}c^{k}%
\end{align}
Here $\left(  \phi_{0},\phi^{m}\right)  $ are bosonic fields and $\left(
\chi_{m},\chi_{0}\right)  $ are fermionic fields. Similarly, we introduce a
gauge parameter in position space $\Lambda\left(  X^{M},c^{m}\right)  $ that
has the expansion%
\begin{equation}
\Lambda\left(  X,c\right)  =\lambda_{0}\left(  X\right)  +c^{m}\Lambda
_{m}\left(  X\right)  +\left(  cc\right)  _{m}\lambda^{m}\left(  X\right)
+\left(  ccc\right)  \Lambda_{0}\left(  X\right)  .
\end{equation}
Here $\left(  \lambda_{0},\lambda^{m}\right)  $ are fermions and $\left(
\Lambda_{m},\Lambda_{0}\right)  $ are bosons. We define the BRST
transformation $\delta_{\Lambda}\Phi=Q\Lambda$ and expand it in powers of $c$
\begin{align}
\delta_{\Lambda}\Phi &  =Q\Lambda=\left(  c^{m}J_{m}-\frac{i}{2}%
\varepsilon^{mnk}c_{m}c_{n}\frac{\partial}{\partial c^{k}}\right)  \left(
\lambda_{0}+c^{m}\Lambda_{m}+\left(  cc\right)  _{m}\lambda^{m}+\left(
ccc\right)  \Lambda_{0}\right) \\
&  =0+c^{m}\left(  J_{m}\lambda_{0}\right)  +\left(  cc\right)  _{m}\left[
\varepsilon^{mnk}J_{n}\Lambda_{k}-i\Lambda^{m}\right]  +\left(  ccc\right)
J_{m}\lambda^{m}\text{. } \label{qL}%
\end{align}
By comparing coefficients we obtain the BRST gauge transformation in terms of
components%
\begin{equation}
\delta_{\Lambda}\phi_{0}=0,\;\delta_{\Lambda}\chi_{m}=J_{m}\lambda
_{0},\;\;\delta_{\Lambda}\phi^{m}=\varepsilon^{mnk}J_{n}\Lambda_{k}%
-i\Lambda^{m},\;\delta_{\Lambda}\chi_{0}=J_{m}\lambda^{m}. \label{qL1}%
\end{equation}
Note that the first component of $\Phi$ is gauge invariant $\delta_{\Lambda
}\phi_{0}=0,$ while the last component of the gauge parameter $\Lambda_{0}$
has no effect on any component of $\Phi.$

Powers of $\Phi$ are easily computed as%
\begin{align}
\Phi^{n}  &  =\phi_{0}^{n}+c^{m}\left[  \left(
%TCIMACRO{\QATOP{n}{1}}%
%BeginExpansion
\genfrac{}{}{0pt}{}{n}{1}%
%EndExpansion
\right)  \phi_{0}^{n-1}\chi_{m}\right]  +\left(  cc\right)  _{m}\left[
\left(
%TCIMACRO{\QATOP{n}{1}}%
%BeginExpansion
\genfrac{}{}{0pt}{}{n}{1}%
%EndExpansion
\right)  \phi_{0}^{n-1}\phi^{m}+\left(
%TCIMACRO{\QATOP{n}{2}}%
%BeginExpansion
\genfrac{}{}{0pt}{}{n}{2}%
%EndExpansion
\right)  \phi_{0}^{n-2}\left(  \chi\chi\right)  ^{m}\right] \\
&  +\left(  ccc\right)  \left[  \left(
%TCIMACRO{\QATOP{n}{1}}%
%BeginExpansion
\genfrac{}{}{0pt}{}{n}{1}%
%EndExpansion
\right)  \phi_{0}^{n-1}\chi_{0}+\left(
%TCIMACRO{\QATOP{n}{2}}%
%BeginExpansion
\genfrac{}{}{0pt}{}{n}{2}%
%EndExpansion
\right)  \phi_{0}^{n-2}\left(  \phi^{m}\chi_{m}\right)  +\left(
%TCIMACRO{\QATOP{n}{3}}%
%BeginExpansion
\genfrac{}{}{0pt}{}{n}{3}%
%EndExpansion
\right)  \phi_{0}^{n-3}\left(  \chi\chi\chi\right)  \right]  .
\end{align}
The gauge transformation properties of these components follow most easily
from Eq.(\ref{qL1}), especially since $\phi_{0}$ is invariant. Hence to
specify the transformation properties of the components of $\delta_{\Lambda
}\Phi^{n}$ we only need the following additional equations that follow from
Eq.(\ref{qL1})
\begin{align}
\delta_{\Lambda}\left(  \phi^{m}\chi_{m}\right)   &  =\left(  \varepsilon
^{mnk}J_{n}\Lambda_{k}-i\Lambda^{m}\right)  \chi_{m}+\phi^{m}J_{m}\lambda
_{0},\\
\delta_{\Lambda}\left(  \chi\chi\right)  ^{m}  &  =\varepsilon^{mnk}\left(
J_{n}\lambda_{0}\right)  \chi_{k}\\
\delta_{\Lambda}\left(  \chi\chi\chi\right)   &  =\frac{1}{2}\varepsilon
^{mnk}\left(  J_{m}\lambda_{0}\right)  \chi_{n}\chi_{k}=\left(  J_{m}%
\lambda_{0}\right)  \left(  \chi\chi\right)  ^{m}.
\end{align}

\subsection{Physical states, BRST cohomology}

The physical states are given by the fields that satisfy the BRST cohomology
\begin{equation}
Q\Phi=0,\;\text{solution identified up to shift }\Phi\rightarrow\Phi+\text{
}Q\Lambda\label{Qcohomology}%
\end{equation}
We will show that these conditions will follow from our proposed action.

We want to demonstrate that the general solution of $Q\Phi=0$ has the form
$\Phi_{solution}=\phi_{0}\left(  X\right)  +Q\Lambda\left(  X,c\right)  ,$
with $J_{m}\phi_{0}\left(  X\right)  =0.$ So the physical field $\Phi$
contains a non-trivial $\phi_{0}\left(  X\right)  ,$ while all other
components described by $Q\Lambda\left(  X,c\right)  $ are considered gauge
freedom, and can be dropped if so desired. The equation $J_{m}\phi_{0}\left(
X\right)  =0$ is equivalent to covariant quantization, interpreted as the
Sp$\left(  2,R\right)  $ gauge invariance condition of the physical field, and
its solution was already discussed in the previous section.

To see that $\Phi\left(  X,c\right)  =\phi_{0}\left(  X\right)  +Q\Lambda
\left(  X,c\right)  ,$ with $J_{m}\phi_{0}\left(  X\right)  =0,$ is the
general solution of the BRST equation $Q\Phi=0$ we examine the components of
the equation by expanding in powers of $c$ as follows
\begin{align}
0 &  =Q\Phi=\left(  c^{m}J_{m}-\frac{i}{2}\varepsilon^{mnk}c_{m}c_{n}%
\frac{\partial}{\partial c^{k}}\right)  \left(  \phi_{0}+c^{m}\chi_{m}+\left(
cc\right)  _{m}\phi^{m}+\left(  ccc\right)  \chi_{0}\right)  \\
&  =0+c^{m}\left(  J_{m}\phi_{0}\right)  +\left(  cc\right)  _{m}\left[
\varepsilon^{mnk}J_{n}\chi_{k}-i\chi^{m}\right]  +\left(  ccc\right)
J_{m}\phi^{m}.\label{qA}%
\end{align}
Therefore the component fields $\left(  \phi_{0},\chi_{m},\phi^{m},\chi
_{0}\right)  \left(  X\right)  $ must satisfy%
\begin{equation}
J_{m}\phi_{0}=0,\;\varepsilon^{mnk}J_{n}\chi_{k}=i\chi^{m},\;J_{m}\phi
^{m}=0.\label{eom}%
\end{equation}
Note that there is no condition on the last component $\chi_{0}.$ The general
solution for $\chi_{m},\varphi^{m},\chi_{0}$ is\footnote{To arrive at this
form first consider the equations (\ref{eom}) as if $J_{m}$ is a classical
vector rather than an operator. Then $J_{m}\varphi^{m}=0$ would be solved by
taking any vector $\varphi^{m}$ perpendicular to $J_{m},$ namely $\varphi
^{m}=\varepsilon^{mnk}J_{n}\Lambda_{k}$ for any vector $\Lambda_{k}.$ However,
since $J_{m}$ is an operator, $J_{m}\varphi^{m}=\varepsilon^{mnk}J_{m}%
J_{n}\Lambda_{k}$ does not vanish, but gives $iJ^{k}\Lambda_{k}$ as a result
of the nonzero Sp$\left(  2,R\right)  $ commutation rules. To compensate for
this effect the solution is modified to $\varphi^{m}=\varepsilon^{mnk}%
J_{n}\Lambda_{k}-i\Lambda^{m},$ and this clearly satisfies $J_{m}\varphi
^{m}=0.$ Similarly, the solution to $\varepsilon^{mnk}J_{n}\chi_{k}=i\chi^{m}$
is $\chi_{m}=J_{m}\lambda_{0}.$ If $J_{m}$ had been a classical vector it
would have given a vanishing result in $\varepsilon^{mnk}J_{n}\chi
_{k}=\varepsilon^{mnk}J_{n}J_{k}\lambda_{0}$, but since it is an Sp$\left(
2,R\right)  $ operator, the result is $\varepsilon^{mnk}J_{n}J_{k}\lambda
_{0}=iJ^{m}\lambda_{0}=i\chi^{m},$ thus satisfying the equation.}%
\begin{equation}
\chi_{m}=J_{m}\lambda_{0},\;\phi^{m}=\varepsilon^{mnk}J_{n}\Lambda
_{k}-i\Lambda^{m},\;\;\chi_{0}\text{=any,}%
\end{equation}
for any $\left(  \lambda_{0},\Lambda^{m},\chi_{0}\right)  .$ We could also
write the arbitrary $\chi_{0}$ in the form $\chi_{0}=J_{m}\lambda^{m}$ for an
arbitrary $\lambda^{m}.$ Not all $\chi_{0}$ can necessarily be written in this
form, so in general the BRST cohomology could be richer in content. However,
we emphasize that our action below has the additional gauge symmetry
$\Phi\rightarrow\Phi+\left(  ccc\right)  \xi\left(  X\right)  $ which is
equivalent to changing $\chi_{0}\left(  X\right)  $ by an arbitrary function
$\xi\left(  X\right)  $. This indicates that in our problem $\chi_{0}$ is pure
gauge freedom, and therefore it can be chosen as we have specified.

By comparing to Eqs.(\ref{qL},\ref{qL1}), we conclude that the components
$\chi_{m},\varphi^{m},\chi_{0}$ of the physical field $\Phi$ can be written as
$Q\Lambda,$ so that the general physical field has the form%
\begin{equation}
\Phi=\phi_{0}+Q\Lambda,\text{ with }J_{m}\phi_{0}=0.
\end{equation}
Hence the BRST cohomology identifies $\phi_{0}$ as the only physical field.
This agrees with the covariant quantization of the 2T-physics theory discussed
in the previous section and in \cite{2tfield}.

We have shown that the quantum version of 2T-physics is correctly reproduced
by the BRST cohomology defined by $Q\Phi=0,$ modulo $\Phi$ of the form
$Q\Lambda$. This combines the differential equations in (\ref{constDiff}) into
a convenient package Q$\Phi=0$ that can be derived from an action principle as
shown below.

\section{Action principle}

\subsection{Free action}

We now propose an action from which we derive the physical state condition
$Q\Phi\left(  X,c\right)  =0$ as an equation of motion. We will also be able
to redefine $\Phi$ by an arbitrary amount $Q\Lambda$ due to a gauge symmetry.
The following form of the free action is inspired from previous work on string
field theory \cite{witten}
\begin{equation}
S\left(  \Phi\right)  =\int\left(  d^{d+2}X\right)  \left(  d^{3}c\right)
\left[  \bar{\Phi}Q\Phi\right]  .
\end{equation}
Here $\bar{\Phi}$ is the hermitian conjugate field. According to the rules of
integration for fermions only the $\left(  ccc\right)  $ term in the product
$\bar{\Phi}Q\Phi$ contributes to the integral. Therefore the action is
obtained in component form by computing the coefficient of $\left(
ccc\right)  $ in $\bar{\Phi}Q\Phi.$ This gives
\begin{equation}
S\left(  \Phi\right)  =\int\left(  d^{d+2}X\right)  \left[  \bar{\phi}%
_{0}J_{m}\phi^{m}+\bar{\phi}^{m}J_{m}\phi_{0}+\varepsilon^{mnk}\bar{\chi}%
_{m}J_{n}\chi_{k}-i\bar{\chi}_{m}\chi^{m}\right]  \label{components1}%
\end{equation}
The action is independent of $\chi_{0}$ because of the gauge symmetry under
the substitution $\Phi\rightarrow\Phi+\left(  ccc\right)  \xi\left(  X\right)
$ with an arbitry $\xi\left(  X\right)  .$ Taking into account the definition
of the $J_{m}$ in terms of $\left(  X,P\right)  ,$ with $P=-i\partial/\partial
X,$ we can do integration by parts. For $J_{0},J_{1}$ integration by parts
just moves the derivatives from $\Phi$ to $\bar{\Phi}$. However, for $J_{2}$
we need to be more careful with signs. Since $J_{2}\Phi=-i\frac{1}{4}\left[
\partial\cdot\left(  X\Phi\right)  +X\cdot\partial\Phi\right]  $ involves a
first order derivative, its integration by parts introduces a minus sign, but
complex conjugation compensates for it and gives the following form
\begin{align}
S\left(  \Phi\right)   &  =\int\left(  d^{d+2}X\right)  \left[  \phi
^{m}\overline{\left(  J_{m}\phi_{0}\right)  }+\phi_{0}\overline{\left(
J_{m}\phi^{m}\right)  }+\varepsilon^{mnk}\chi_{m}\overline{\left(  J_{n}%
\chi_{k}\right)  }+i\chi^{m}\bar{\chi}_{m}\right]  \label{components2}\\
&  =\int\left(  d^{d+2}X\right)  \left(  d^{3}c\right)  \left[  \Phi
\overline{\left(  Q\Phi\right)  }\right]
\end{align}
where $\overline{\left(  J_{m}\phi_{0}\left(  X\right)  \right)  }$ is the
complex conjugate after applying the differential operators $J_{m}$ on
$\phi_{0},$ etc. In obtaining this result we have also inserted extra minus
signs in changing orders of two fermions such as $-i\bar{\chi}_{m}\chi^{m}=$
$+i\chi^{m}\bar{\chi}_{m}$. From this result on integration by parts, we see
that the latter form of the Lagrangian $\Phi\overline{\left(  Q\Phi\right)  }$
is the hermitian conjugate of the original expression $\bar{\Phi}Q\Phi$.
Furthermore we can apply the same method to argue that we can perform
integration by parts for any two fields as follows
\begin{equation}
\int\bar{A}QB=\left(  -1\right)  ^{AB+A+B}\int B\overline{\left(  QA\right)
}.
\end{equation}
where the sign factor takes into account the Grassmann parity $\left(
-1\right)  ^{A}$ or $\left(  -1\right)  ^{B}$ of the fields $A,B.$

Using these properties of the integral we can now study the variation of the
action%
\begin{align}
\delta S  &  =\int\left(  d^{d+2}X\right)  \left(  d^{3}c\right)  \left[
\overline{\left(  \delta\Phi\right)  }Q\Phi+\bar{\Phi}Q\left(  \delta
\Phi\right)  \right] \\
&  =\int\left(  d^{d+2}X\right)  \left(  d^{3}c\right)  \left[  \overline
{\left(  \delta\Phi\right)  }Q\Phi+\delta\Phi\overline{\left(  Q\Phi\right)
}\right]
\end{align}
where in the second term we have used the property of integration by parts.
From this we see that the action yields the desired equations of motion for
the general variation $\delta\Phi$
\begin{equation}
Q\Phi=0,
\end{equation}
and its hermitian conjugate. Furthermore we can also verify that the action
has the gauge symmetry under the special transformation $\delta_{\Lambda}%
\Phi=Q\Lambda$%
\begin{align}
\delta_{\Lambda}S  &  =\int\left(  d^{d+2}X\right)  \left(  d^{3}c\right)
\left[  \overline{\left(  \delta_{\Lambda}\Phi\right)  }Q\Phi+\left(
\delta_{\Lambda}\Phi\right)  \overline{\left(  Q\Phi\right)  }\right] \\
&  =\int\left(  d^{d+2}X\right)  \left(  d^{3}c\right)  \left[  \overline
{\left(  Q\Lambda\right)  }Q\Phi+\left(  Q\Lambda\right)  \overline{\left(
Q\Phi\right)  }\right] \\
&  =\int\left(  d^{d+2}X\right)  \left(  d^{3}c\right)  \left[  \overline
{\Lambda}Q^{2}\Phi+\left(  Q^{2}\Lambda\right)  \overline{\Phi}\right]  =0.
\end{align}
In the third line we have used again the rule for integration by parts and
applied the property $Q^{2}=0.$ This justifies that the physical state must be
taken only up to the gauge transformations. Hence only the $\phi_{0}$
component of $\Phi$ contains the physical field since the remainder can be
gauge fixed even off-shell. This is consistent with the BRST cohomology of
Eq.(\ref{Qcohomology}) that we discussed in the previous section, but now we
have derived it from an action principle.

It is possible to verify these properties directly in terms of components by
using Eqs.(\ref{components1},\ref{components2}), and we leave this as an
exercise for the reader.

\subsection{Interactions in equation of motion form}

Given the analogy to string field theory, one may be tempted to consider the
Cherns-Simons type action by taking a field $\left(  A\left(  X,c\right)
\right)  _{i}^{~j}$ that is also a matrix. The expansion in powers of $c^{m}$
has matrix coefficients $A\left(  X,c\right)  =a_{0}\left(  X\right)
+c^{m}\xi_{m}\left(  X\right)  +\left(  cc\right)  _{m}a^{m}\left(  X\right)
+\left(  ccc\right)  \zeta_{0}\left(  X\right)  ,$ and the interaction for a
Chern-Simons action is%
\begin{equation}
S_{cs}\left(  A\right)  =\int\left(  d^{d+2}X\right)  \left(  d^{3}c\right)
\text{ Tr}\left[  \frac{1}{2}AQA+\frac{1}{3}AAA\right]  .
\end{equation}
Unfortunately, this cannot be a valid action for a hermitian $A\left(
X,c\right)  .$ The difficulty is rooted in the integration by parts property
for $J_{2}$ that we discussed following Eq.(\ref{components1}). For a
hermitian field the terms involving $J_{2}$ drop out after an integration by
parts, and therefore the action above cannot be a correct starting point. This
is just as well, since we were in the process of discussing a scalar field
$\Phi\left(  X,c\right)  $ that reduced to the Klein-Gordon field
$\varphi\left(  x\right)  $, while the Chern-Simons type action usually is
expected to lead to a vector gauge field $\left(  A_{\mu}\left(  x\right)
\right)  _{i}^{~j}$. We expect to return to to the Chern-Simons type action
when we discuss spinning particles including a gauge vector field.

So, for the present paper we concentrate on interactions of $\Phi\left(
X,c\right)  $ that will lead to the local self interactions of the scalar
field $\varphi\left(  x\right)  $ in $\left(  d-1\right)  +1$ dimensions. The
action must still maintain a gauge symmetry of the type $\delta_{\Lambda}%
\Phi=Q\Lambda+\cdots$ in the presence of interactions in order to remove the
unphysical degrees of freedom $\phi^{m},\chi_{m},\chi_{0}$. Here the $\left(
+\cdots\right)  $ is a possible modification of the transformation laws due to
the interaction.

We are aiming for the following form of interaction that we know from old work
\cite{2tfield} to be consistent at the equation of motion level
\begin{align}
Q_{11}\phi_{0}  &  =\frac{X^{2}}{2}\phi_{0},\;Q_{12}\phi_{0}=\frac
{X\cdot\partial_{X}+\partial_{X}\cdot X}{4i}\phi_{0},\;\\
Q_{22}\phi_{0}  &  =\frac{1}{2}\left(  -\partial_{X}^{2}\phi_{0}+\Delta\left(
\bar{\phi}_{0}\phi_{0}\right)  \phi_{0}\right)  ,\;\text{where }\Delta\left(
\bar{\phi}_{0}\phi_{0}\right)  =-\tilde{\lambda}\left(  \bar{\phi}_{0}\phi
_{0}\right)  ^{2/\left(  d-2\right)  }.
\end{align}
Only $Q_{22}$ is modified by including the interaction. These interacting
$Q_{ij}$ satisfy the Sp$\left(  2,R\right)  $ commutation rules
\begin{equation}
\left[  Q_{11},Q_{22}\right]  \phi_{0}=2iQ_{12}\phi_{0},\;\;\left[
Q_{12},Q_{11}\right]  \phi_{0}=-iQ_{11}\phi_{0},\;\;\left[  Q_{12}%
,Q_{22}\right]  \phi_{0}=iQ_{22}\phi_{0}, \label{consistency}%
\end{equation}
provided $\phi_{0}$ is on shell $Q_{ij}\phi_{0}=0$ (actually only $Q_{11}%
\phi_{0}=Q_{12}\phi_{0}=0$ is enough). Therefore, the following equations of
motion can be applied consistently including a unique interaction term, with
$\Delta\left(  \bar{\phi}_{0}\phi_{0}\right)  $ given above,
\begin{equation}
Q_{11}\phi_{0}=0,\;\;Q_{12}\phi_{0}=0,\;\;Q_{22}\phi_{0}=0.
\end{equation}
The uniqueness of the interaction is determined by the consistency conditions
of Eq.(\ref{consistency}) as follows.

The solution of $Q_{11}\phi_{0}=0$ is still $\phi_{0}=\delta\left(
X^{2}\right)  A\left(  X\right)  $ as before (section \ref{cov}). Requiring
this to satisfy $Q_{12}\phi_{0}=0$ is equivalent to demanding $A$ to have a
definite dimension, again same as before (section \ref{cov})
\begin{equation}
X\cdot\partial_{X}A=-\frac{d-2}{2}A.
\end{equation}
To check the consistency of the interacting case as in Eq.(\ref{consistency})
we also need to compute $Q_{12}$ applied on functionals of $\phi_{0},$ such as
powers
\begin{equation}
\phi_{0}^{n}=\left(  \delta\left(  X^{2}\right)  A\left(  X\right)  \right)
^{n}=\frac{\delta\left(  X^{2}\right)  }{\delta\left(  0\right)  }\left(
\delta\left(  0\right)  A\left(  X\right)  \right)  ^{n}=\sigma^{n-1}%
\delta\left(  X^{2}\right)  \left(  A\left(  X\right)  \right)  ^{n},
\end{equation}
where for convenience\footnote{As explained following Eq.(\ref{renCoupling}),
after absorbing the factors of $\sigma,$ the renormalized coupling
$\lambda=\tilde{\lambda}\sigma^{4/(d-2)}$ that appears in the interaction
terms among the fields $A\left(  X\right)  $ or $\varphi\left(  x\right)  $ is
finite, so we do not need to be concerned with the value of $\sigma
.\label{renormalized}$} we have defined $\sigma\equiv\delta\left(  0\right)
.$ Such powers $\phi_{0}^{n}$ have anomalous dimensions that shift the
dimension away from the naive dimension, namely $X\cdot\partial_{X}\phi
_{0}^{n}\neq-n\frac{d-2}{2}\phi_{0}^{n}.$ This is because the $X$ dependence
of the delta functions have disappeared into the constant $\sigma$ except for
one factor. This is discussed in more detail in section (\ref{anom}). Having
noted this, the action of $Q_{12}$ is computed as follows
\begin{align}
Q_{12}\left(  \frac{\phi_{0}^{n}}{\sigma^{n-1}}\right)   &  =\left[
\frac{X\cdot\partial_{X}}{2i}\delta\left(  X^{2}\right)  \right]  A^{n}%
+\delta\left(  X^{2}\right)  A^{n-1}\left[  n\frac{X\cdot\partial_{X}}%
{2i}A\right]  +\frac{d+2}{4i}\delta\left(  X^{2}\right)  A^{n}\\
&  =i\delta\left(  X^{2}\right)  A^{n}+\delta\left(  X^{2}\right)  A^{n}%
\frac{-n\left(  d-2\right)  }{4i}+\frac{d+2}{4i}\delta\left(  X^{2}\right)
A^{n}\\
&  =i\left(  \frac{\phi_{0}^{n}}{\sigma^{n-1}}\right)  \left(  1+\frac
{n\left(  d-2\right)  }{4}-\frac{d+2}{4}\right) \\
&  =i\frac{\left(  n-1\right)  \left(  d-2\right)  }{4}\left(  \frac{\phi
_{0}^{n}}{\sigma^{n-1}}\right)  .
\end{align}
This formula is used to verify that the commutation rule $\left[
Q_{12},Q_{22}\right]  \phi_{0}=iQ_{22}\phi_{0}$ works only when%
\begin{equation}
\Delta=-\tilde{\lambda}\left(  \bar{\phi}_{0}\phi_{0}\right)  ^{\frac{2}{d-2}%
}.
\end{equation}

We have shown that the equations of motion $Q_{11}\phi_{0}=0,\;Q_{12}\phi
_{0}=0,\;Q_{22}\phi_{0}=0,$ become%
\begin{equation}
\left(  X\cdot\frac{\partial A}{\partial X}+\frac{d-2}{2}A\right)  _{X^{2}%
=0}=0,\;\;\left(  -\frac{\partial^{2}A}{\partial X_{M}\partial X^{M}}%
-\lambda\left(  \bar{A}A\right)  ^{\frac{2}{d-2}}A\right)  _{X^{2}=0}=0.
\end{equation}
The $\lambda$ here is the renormalized constant$^{\ref{renormalized}}.$ These
were the interacting equations obtained before in \cite{2tfield}, and now we
have written them in the form $Q_{ij}\phi_{0}=0$ for $\phi_{0}$ instead of
$A.$ In terms of $\phi_{0}$ they have the BRST form, including interactions,
so we may expect to derive them from a BRST type action by generalizing the
free BRST action to an interacting one, as will be done below.

\subsection{Interacting action and gauge symmetry \label{symmetry}}

The action is obtained by starting with the free BRST action and then
modifying $Q_{ij}$ as above. It is convenient to rewrite the field components
in the basis $Q_{ij}$ of Eq.(\ref{Qij}) instead of the basis of the $J_{m}$ of
Eq.(\ref{Jm}). Including only $\phi_{0}$ and $\phi^{m}~$($=\phi_{11},\phi
_{12},\phi_{22},$ in the new notation), the free action in
Eq.(\ref{components1}) is generalized to the following interacting action
(suppressing the fermionic components $\chi^{m},\chi_{0}$ that decouple)%
\begin{equation}
S=\int\left(  d^{d+2}X\right)  \left\{  \left[  \bar{\phi}_{22}Q_{11}^{0}%
\phi_{0}-2\bar{\phi}_{12}Q_{12}^{0}\phi_{0}+\bar{\phi}_{11}\left(  Q_{22}%
^{0}+\Delta\right)  \phi_{0}+h.c.\right]  -U\left(  \bar{\phi}_{0}\phi
_{0}\right)  \right\}  . \label{action}%
\end{equation}
We included the superscript \textquotedblleft0\textquotedblright\ to emphasize
that $Q_{ij}^{0}$ are the free Sp$\left(  2,R\right)  $ generators of
Eq.(\ref{Qij},\ref{constDiff}). There are two sources of interactions. One is
\begin{equation}
\Delta=-\tilde{\lambda}\left(  \bar{\phi}_{0}\phi_{0}\right)  ^{2/\left(
d-2\right)  }, \label{D}%
\end{equation}
which is part of $Q_{22}=Q_{22}^{0}+\Delta$ as motivated in the previous
section by studying the on-shell equations of motion. The second one,
$U\left(  \bar{\phi}_{0}\phi_{0}\right)  ,$ is like a potential energy term
given below. We can actually attempt to take an arbitrary function
$\Delta\left(  \bar{\phi}_{0}\phi_{0}\right)  $ rather than the specific one
given above. However, after going through all the calculations, we can show
that only the specific form of $\Delta\left(  \bar{\phi}_{0}\phi_{0}\right)  $
above is consistent with all the symmetries generated by the SL$\left(
2,R\right)  $ generators. Therefore we start from the beginning with the form
given in Eq.(\ref{D}) to simplify our presentation.

If the gauge transformations were the free ones of Eq.(\ref{qL1}) then any
potential term $U\left(  \bar{\phi}_{0}\phi_{0}\right)  $ would be allowed
since $\phi_{0}$ would be gauge invariant. However, although $\phi_{0}$ is
gauge invariant on shell in the interacting theory (since $Q_{ij}\phi_{0}=0$
with the full $Q_{ij}$), it is not gauge invariant off shell $Q_{ij}\phi
_{0}\neq0$. Then the off-shell gauge symmetry discussed below, together with
the on-shell consistency of Eqs.(\ref{consistency}), require that the form of
$U$ should be restricted, such that $\Delta$ and $U$ are related by
$\frac{\partial U}{\partial\bar{\phi}_{0}}\sim\phi_{0}\Delta$ up to a
proportionality constant that will be determined below consistently. Given
$\Delta$ above, we must have then%
\begin{equation}
U=a\tilde{\lambda}\left(  \bar{\phi}_{0}\phi_{0}\right)  ^{d/\left(
d-2\right)  }, \label{W}%
\end{equation}
where $a$ is a constant whose significance will be discussed below.

We now discuss the modified gauge symmetry in the presence of interactions.
Since the modified $Q_{ij}$ close when $\phi_{0}$ is on shell as in
Eq.(\ref{consistency}), we might expect that the action has an
\textit{off-shell} gauge symmetry generated by the interacting $Q_{ij}$ with
some appropriate modification of the transformation rules given in
Eq.(\ref{qL1}). Indeed we have found that the interacting action is invariant
under modified gauge transformations which are obtained as follows

\begin{itemize}
\item Substitute the interacting $Q_{ij}$ instead of the free $Q_{ij}^{0}$ in
the transformation rules of $\delta_{\Lambda}\phi_{ij}.$

\item Include an additional modification denoted by $\gamma$ in the last term
of $\delta_{\Lambda}\phi_{12}$ as shown in Eqs.(\ref{t4},\ref{gamma}) below.
\end{itemize}

Including the modifications just noted, the transformation rules of
Eq.(\ref{qL1}) for $\phi^{m}$ get generalized from the free case $\delta
\phi_{m}=\varepsilon_{mnk}J^{n}\Lambda^{k}-i\Lambda_{m}$ to the interacting
case by the inclusion of $\Delta$ as follows (with change of notation
$\phi_{11}=\phi_{0}-\phi_{1},\;\phi_{22}=\phi_{0}+\phi_{1},\;\phi_{12}%
=\phi_{2}$)
\begin{align}
\delta_{\Lambda}\phi_{11}  &  =Q_{11}^{0}\Lambda_{12}-Q_{12}^{0}\Lambda
_{11}-i\Lambda_{11},\;\;\label{t1}\\
\delta_{\Lambda}\phi_{22}  &  =Q_{12}^{0}\Lambda_{22}-\left(  Q_{22}%
^{0}+\Delta\right)  \Lambda_{12}-i\Lambda_{22},\label{t2}\\
\delta_{\Lambda}\phi_{12}  &  =\frac{1}{2}Q_{11}^{0}\Lambda_{22}-\frac{1}%
{2}\left(  Q_{22}^{0}+\Delta\right)  \Lambda_{11}-i\left(  \Lambda_{12}%
+\gamma\right)  \label{t3}%
\end{align}
where we insert
\begin{equation}
\gamma=\frac{i\Delta}{d-2}\left(  \Lambda_{11}-\frac{\phi_{0}}{\bar{\phi}_{0}%
}\bar{\Lambda}_{11}\right)  . \label{gamma}%
\end{equation}
Let's prove this gauge symmetry of the action. The gauge transformation of the
action gives%
\begin{equation}
\delta_{\Lambda}S=\int\left(  d^{d+2}X\right)  \left[
\begin{array}
[c]{c}%
\left(  Q_{12}^{0}\Lambda_{22}-\left(  Q_{22}^{0}+\Delta\right)  \Lambda
_{12}-i\Lambda_{22}\right)  ^{\ast}Q_{11}^{0}\phi_{0}\\
-2\left(  \frac{1}{2}Q_{11}^{0}\Lambda_{22}-\frac{1}{2}\left(  Q_{22}%
^{0}+\Delta\right)  \Lambda_{11}-i\left(  \Lambda_{12}+\gamma\right)  \right)
^{\ast}Q_{12}^{0}\phi_{0}\\
+\left(  Q_{11}^{0}\Lambda_{12}-Q_{12}^{0}\Lambda_{11}-i\Lambda_{11}\right)
^{\ast}\left(  Q_{22}^{0}+\Delta\right)  \phi_{0}+h.c.
\end{array}
\right]
\end{equation}
After integrating by parts and collecting coefficients of the $\Lambda_{ij},$
one finds that most terms cancel by using the free commutation rules $\left[
Q_{11}^{0},Q_{22}^{0}\right]  =2iQ_{12}^{0}$,\ \ $\left[  Q_{12}^{0}%
,Q_{11}^{0}\right]  =-iQ_{11}^{0}$,\ \ $\left[  Q_{12}^{0},Q_{22}^{0}\right]
=iQ_{22}^{0}$, as well as $\Delta Q_{11}^{0}-Q_{11}^{0}\Delta=0$ (since
$Q_{11}^{0}=\frac{1}{2}X^{2}$ is not a differential operator). The remaining
terms involve $\Delta$ and $\gamma$ as follows%
\begin{equation}
\delta_{\Lambda}S=\int\left(  d^{d+2}X\right)  \left[  \left\{
\begin{array}
[c]{c}%
\bar{\Lambda}_{11}\left[  \Delta Q_{12}^{0}\phi_{0}-Q_{12}^{0}\left(
\Delta\phi_{0}\right)  +i\Delta\phi_{0}\right]  +h.c.\\
-2i\gamma Q_{12}^{0}\phi_{0}-\left(  2i\gamma Q_{12}^{0}\phi_{0}\right)
^{\ast}.
\end{array}
\right\}  \right]  \label{Qf}%
\end{equation}
We are aiming to show that the $\gamma,\gamma^{\ast}$ terms cancel the
$\Delta$ terms, so we study the $\Delta$ terms in more detail. In this
expression $Q_{12}^{0}$ is the following derivative operator as applied on any
function of $X^{M}$ in $d+2$ dimensions
\begin{align}
Q_{12}^{0}f\left(  X\right)   &  =-\frac{i}{4}\left(  X\cdot\partial
_{X}+\partial_{X}\cdot X\right)  f\left(  X\right) \\
&  =-\frac{i}{2}X\cdot\partial_{X}f\left(  X\right)  -\frac{i}{4}\left(
d+2\right)  f\left(  X\right)  .
\end{align}
If $\phi_{0}\left(  X\right)  $ were a smooth function, applying this operator
in the first line of Eq.(\ref{Qf}), and using the chain rule for derivatives,
we would obtain the result
\begin{align}
&  \Delta Q_{12}^{0}\phi_{0}-Q_{12}^{0}\left(  \Delta\phi_{0}\right)
+i\Delta\phi_{0}\\
&  =\frac{i}{2}\left(  \phi_{0}X\cdot\partial_{X}\Delta+2\phi_{0}\Delta\right)
\label{wrong1}\\
&  =\frac{i}{2}\frac{2\Delta}{d-2}\left[  X\cdot\partial\phi_{0}+\frac
{\phi_{0}}{\bar{\phi}_{0}}X\cdot\partial\bar{\phi}_{0}+\left(  d-2\right)
\phi_{0}\right]  \label{wrong}%
\end{align}
where the explicit form of $\Delta$ in Eq.(\ref{D}) is used to compute the
derivative $X\cdot\partial_{X}\Delta.$

However, in computing $Q_{12}^{0}\left(  \Delta\phi_{0}\right)  $ we must be
careful about the fact that $\phi_{0}\left(  X\right)  $ is very singular.
Recall that one of the equations of motion $0=Q_{11}^{0}\phi_{0}=\frac{1}%
{2}X^{2}\phi_{0},$ requires the singular configuration $\phi_{0}\left(
X\right)  =\delta\left(  X^{2}\right)  A\left(  X\right)  .$ Therefore,
functionals such as $\Delta\left(  \bar{\phi}_{0}\phi_{0}\right)  \phi_{0}$
that appear in Eq.(\ref{Qf}) have to be differentiated carefully by taking
into account the singular nature. As we will show in the following subsection,
this has the effect of producing an anomalous dimension for the functional
$\Delta\left(  \bar{\phi}_{0}\phi_{0}\right)  \phi_{0}$, such that, instead of
the $\left(  d-2\right)  \phi_{0}$ in the last term of Eq.(\ref{wrong}) we
obtain $\left(  d+2\right)  \phi_{0}$. After including the anomalous dimension
(see second line of equation below), the $\Delta$ terms in $\delta_{\Lambda}S$
in Eq.(\ref{Qf}) become
\begin{align}
&  \bar{\Lambda}_{11}\left[  \Delta Q_{12}^{0}\phi_{0}-Q_{12}^{0}\left(
\Delta\phi_{0}\right)  +i\Delta\phi_{0}\right] \\
&  =\frac{i}{2}\frac{2\bar{\Lambda}_{11}\phi_{0}\Delta}{d-2}\left[
X\cdot\partial\ln\phi_{0}+X\cdot\partial\ln\bar{\phi}_{0}+\left(  d+2\right)
\right]  \label{correct}%
\end{align}
Two significant features of this expression are: first, it is proportional to
$\phi_{0}\Delta$ which can be obtained as the derivative of the function
$U=a\tilde{\lambda}\left(  \bar{\phi}_{0}\phi_{0}\right)  ^{d/\left(
d-2\right)  }$ up to a proportionality constant, and second, it vanishes on
shell because it is proportional to $Q_{12}^{0}\phi_{0},\left(  Q_{12}^{0}%
\phi_{0}\right)  ^{\ast}$ as shown below
\begin{align}
&  X\cdot\partial\ln\phi_{0}+X\cdot\partial\ln\bar{\phi}_{0}+\left(
d+2\right) \\
&  =\frac{2i}{\phi_{0}}\left(  \frac{1}{2i}X\cdot\partial\phi_{0}+\frac{1}%
{4i}\left(  d+2\right)  \phi_{0}\right)  +\frac{2i}{\bar{\phi}_{0}}\left(
\frac{1}{2i}X\cdot\partial\bar{\phi}_{0}+\frac{1}{4i}\left(  d+2\right)
\bar{\phi}_{0}\right) \\
&  =2i\left[  \frac{1}{\phi_{0}}Q_{12}^{0}\phi_{0}-\left(  \frac{1}{\phi_{0}%
}Q_{12}^{0}\phi_{0}\right)  ^{\ast}\right]
\end{align}
Including the hermitian conjugate the $\Delta$ terms can be rewritten in the
following form
\begin{align}
&  \bar{\Lambda}_{11}\left[  \Delta Q_{12}^{0}\phi_{0}-Q_{12}^{0}\left(
\Delta\phi_{0}\right)  +i\Delta\phi_{0}\right]  +h.c.\\
&  =-\frac{2\Delta}{d-2}\left[  \left(  \bar{\Lambda}_{11}-\frac{\phi_{0}%
}{\bar{\phi}_{0}}\Lambda_{11}\right)  Q_{12}^{0}\phi_{0}\right]  +h.c.
\end{align}
Therefore we can cancel the $\Delta$ terms in $\delta_{\Lambda}S$ in
Eq.(\ref{Qf}) by taking $\gamma$ as given in Eq.(\ref{gamma}). With this
choice Eq.(\ref{Qf}) becomes%
\[
\delta_{\Lambda}S\;=0,
\]
off shell.

We have constructed a gauge invariant interacting 2T field theory in $d+2$
dimensions, with the gauge transformation rules for $\left(  \phi_{11}%
,\phi_{12},\phi_{22}\right)  $ given by Eqs.(\ref{t1}-\ref{t3}), while
$\gamma$ is explicitly given in (\ref{gamma}). The gauge symmetry compensates
for one extra space and one extra time dimensions, making this field theory
free of ghosts, and closely related to 1T-physics field theory in $\left(
d-1\right)  +1$ dimensions. There are many ways of descending from $d+2$
dimensions to $\left(  d-1\right)  +1$ dimensions as illustrated in Fig.1.
Therefore, the 2T-physics field theory we have constructed is expected to lead
to a variety of 1T-physics field theories, each being a holographic image of
the parent theory, and having duality type relations among themselves. The
duality transformation is related to the gauge transformations we discussed in
this section.

\subsection{Anomalous dimension \label{anom}}

Keeping in mind the on-shell singular behavior $\phi_{0}=\delta\left(
X^{2}\right)  A\left(  X\right)  ,$ we will assume that the off shell
$\phi_{0} $ that is relevant in our theory must be similarly singular. So, we
will assume that the anomalous dimension of the off-shell expression
$\Delta\left(  \bar{\phi}_{0}\phi_{0}\right)  \phi_{0}$ is the same as the one
on-shell, so we compute the anomalous dimension for the on-shell quantity.
First note that the powers of the delta function $\left(  \delta\left(
X^{2}\right)  \right)  ^{n}$ collapse to a single overall delta function while
the rest are evaluated at $\delta\left(  0\right)  .$ We will define the
constant $\delta\left(  0\right)  \equiv\sigma$ for brevity. Then we can
write
\begin{equation}
\Delta\left(  \bar{\phi}_{0}\phi_{0}\right)  \phi_{0}=\delta\left(
X^{2}\right)  ~A~\Delta\left(  \sigma^{2}\bar{A}A\right)  .
\end{equation}
For $\Delta$ as given in Eq.(\ref{D}) the constant $\sigma$ can be absorbed
away into a redefinition of the coupling constant $\tilde{\lambda}$, so we do
not need to be concerned about the infinite value of $\sigma$ and simply treat
it formally as a constant in the following arguments. Now we compute the first
derivative $X^{M}\frac{\partial}{\partial X^{M}}$ of this expression (i.e.
dimension operator applied on $\Delta\left(  \bar{\phi}_{0}\phi_{0}\right)
\phi_{0}$)
\begin{align}
&  X\cdot\partial\left(  \Delta\left(  \bar{\phi}_{0}\phi_{0}\right)  \phi
_{0}\right) \\
&  =X\cdot\partial\left[  \delta\left(  X^{2}\right)  ~A~\Delta\left(
\sigma^{2}\bar{A}A\right)  \right] \\
&  =A\delta\left(  X^{2}\right)  ~X\cdot\partial\Delta\left(  \sigma^{2}%
\bar{A}A\right)  +\Delta\left(  \sigma^{2}\bar{A}A\right)  ~X\cdot
\partial\left(  A\delta\left(  X^{2}\right)  \right)  \label{preanom}%
\end{align}
In the last term we can use $X\cdot\partial\delta\left(  X^{2}\right)
=-2\delta\left(  X^{2}\right)  $ as in Eq.(\ref{delta2}) to argue that every
term in this expression is proportional to $\delta\left(  X^{2}\right)  .$
Note that in evaluating $X\cdot\partial\Delta\left(  \sigma^{2}\bar
{A}A\right)  $ in what follows, $\sigma=\delta\left(  0\right)  $ will not
contribute a similar factor to $X\cdot\partial\delta\left(  X^{2}\right)
=-2\delta\left(  X^{2}\right)  $, and this is the subtlety that leads to an
anomalous dimension. With this understanding we proceed with the following
computation by using the explicit form of $\Delta$ given in Eq.(\ref{D})%
\begin{align}
&  \delta\left(  X^{2}\right)  ~A~X\cdot\partial\Delta\left(  \sigma^{2}%
\bar{A}A\right) \\
&  =\frac{2}{d-2}\delta\left(  X^{2}\right)  ~A~\Delta\left(  \sigma^{2}%
\bar{A}A\right)  ~X\cdot\partial\ln\left(  \bar{A}A\right) \\
&  =\frac{2}{d-2}\delta\left(  X^{2}\right)  ~A~\Delta\left(  \sigma^{2}%
\bar{A}A\right)  ~\left[  \frac{1}{A}X\cdot\partial A+\frac{1}{\bar{A}}%
X\cdot\partial\bar{A}\right] \\
&  =\frac{2}{d-2}\Delta\left(  \sigma^{2}\bar{A}A\right)  ~\left[
\delta\left(  X^{2}\right)  ~X\cdot\partial A+~\frac{A}{\bar{A}}~\delta\left(
X^{2}\right)  ~X\cdot\partial\bar{A}\right] \\
&  =\frac{2}{d-2}\Delta\left(  \sigma^{2}\bar{A}A\right)  \left[
X\cdot\partial\left(  A\delta\left(  X^{2}\right)  \right)  +\frac{A}{\bar{A}%
}X\cdot\partial\left(  \bar{A}\delta\left(  X^{2}\right)  \right)
+4A\delta\left(  X^{2}\right)  \right]
\end{align}
In the last line we pulled $\delta\left(  X^{2}\right)  $ inside the
derivative by using again $X\cdot\partial\delta\left(  X^{2}\right)
=-2\delta\left(  X^{2}\right)  $ as in Eq.(\ref{delta2}). Now we combine this
result with Eq.(\ref{preanom}) and obtain%
\begin{align}
&  X\cdot\partial\left[  A\delta\left(  X^{2}\right)  ~\Delta\left(
\sigma^{2}\bar{A}A\right)  \right] \\
&  =\frac{2}{d-2}\Delta\left(  \sigma^{2}\bar{A}A\right)  \left[
X\cdot\partial\left(  A\delta\left(  X^{2}\right)  \right)  +4A\delta\left(
X^{2}\right)  \right]  +\Delta\left(  \sigma^{2}\bar{A}A\right)
~X\cdot\partial\left(  A\delta\left(  X^{2}\right)  \right) \\
&  +\frac{2}{d-2}\frac{A}{\bar{A}}\Delta\left(  \sigma^{2}\bar{A}A\right)
X\cdot\partial\left(  \bar{A}\delta\left(  X^{2}\right)  \right) \\
&  =\frac{2}{d-2}\Delta\left(  \bar{\phi}_{0}\phi_{0}\right)  \left[
X\cdot\partial\phi_{0}+\frac{\phi_{0}}{\bar{\phi}_{0}}X\cdot\partial\bar{\phi
}_{0}+4\phi_{0}\right]  +\Delta\left(  \bar{\phi}_{0}\phi_{0}\right)
X\cdot\partial\phi_{0}%
\end{align}
In the final line we replaced back $\phi_{0}=A\left(  X\right)  \delta\left(
X^{2}\right)  .$ In this way we have derived%
\begin{align}
X\cdot\partial\left(  \Delta\left(  \bar{\phi}_{0}\phi_{0}\right)  \phi
_{0}\right)   &  =\frac{2}{d-2}\Delta\left(  \bar{\phi}_{0}\phi_{0}\right)
\left[  X\cdot\partial\phi_{0}+\frac{\phi_{0}}{\bar{\phi}_{0}}X\cdot
\partial\bar{\phi}_{0}+4\phi_{0}\right] \\
&  +\Delta\left(  \bar{\phi}_{0}\phi_{0}\right)  X\cdot\partial\phi_{0}%
\end{align}
The term $4\phi_{0}$ inside the bracket has no derivatives, so this is the
anomalous term that would not arise if we had smooth functions. When this
$4\phi_{0}$ is combined with the $\left(  d-2\right)  \phi_{0}$ in
Eq.(\ref{wrong}) it leads to $\left(  d+2\right)  \phi_{0}$ that appears in
Eq.(\ref{correct}).

We computed the anomalous dimension for the on-shell quantity $\Delta\left(
\bar{\phi}_{0}\phi_{0}\right)  \phi_{0}$ by imposing only one of the on-shell
conditions, while the other two conditions are still off-shell. At this stage
we do not have a corresponding proof for the fully off-shell quantity, but
assumed that the same result holds. In any case it is evident that for field
configurations that are singular like $\delta\left(  X^{2}\right)  $ the
result holds, and for those field configurations the gauge invariance of the
action is valid.

\section{Standard 1T-field theory from 2T-field theory}

Having justified the gauge symmetry, we can now make various gauge choices for
$\phi_{ij}$. We would like to show that in a specific gauge the 2T-physics
field theory in $d+2$ dimensions descends to the standard interacting
Klein-Gordon field theory in $\left(  d-1\right)  +1$ dimensions. We emphasize
that, in other gauges as in Fig.1, we expect to find other field theories
instead of the Klein-Gordon theory, so the discussion below is an illustration
of the reduction from 2T-physics to 1T-physics in the context of field theory.
This reduction is done in this section at the level of the action and should
be compared to the corresponding reduction at the level of equations of motion
discussed earlier in this paper and in \cite{2tfield}.

The 2T action in Eq.(\ref{action}) contains the fields $\phi_{ij}$ and
$\phi_{0}.$ We must require that the equations of motion that follow from the
original action, after gauge fixing, agree with the equations of motion
produced by the gauge fixed action. So we start by writing down all equations
of motion and then fix the gauges. The variation of the action with respect to
$\delta\bar{\phi}_{11}$ before choosing a gauge gives the equation%
\begin{equation}
Q_{22}^{0}\phi_{0}+\phi_{0}\Delta=0,\;\text{with }\Delta=-\tilde{\lambda
}\left(  \bar{\phi}_{0}\phi_{0}\right)  ^{2/\left(  d-2\right)  }.
\label{f0eqs}%
\end{equation}
The variation $\delta\bar{\phi}_{0}$ before choosing a gauge gives the
equation\footnote{Observe that, including the hermitian conjugate, the
interaction with $\phi_{11}$ is of the form $-\tilde{\lambda}\bar{\phi}%
_{11}\left(  \bar{\phi}_{0}\right)  ^{\frac{2}{d-2}}\left(  \phi_{0}\right)
^{\frac{d}{d-2}}-\tilde{\lambda}\left(  \bar{\phi}_{0}\right)  ^{\frac{d}%
{d-2}}\left(  \phi_{0}\right)  ^{\frac{2}{d-2}}\phi_{11}.$}
\begin{equation}
Q_{11}^{0}\phi_{22}-2Q_{12}^{0}\phi_{12}=-Q_{22}^{0}\phi_{11}-\left[
(\frac{\phi_{11}d}{\phi_{0}}+\frac{2\bar{\phi}_{11}}{\bar{\phi}_{0}%
})+ad\right]  \frac{\phi_{0}\Delta}{d-2}. \label{feq}%
\end{equation}
where we have used $\frac{\partial U}{\partial\bar{\phi}_{0}}=\frac
{ad\tilde{\lambda}}{d-2}\phi_{0}\left(  \bar{\phi}_{0}\phi_{0}\right)
^{2/\left(  d-2\right)  }=-\frac{ad}{d-2}\phi_{0}\Delta$.

Now we choose the gauge
\begin{equation}
\phi_{11}=\gamma\phi_{0}, \label{gauge11}%
\end{equation}
where $\gamma$ is a constant real coefficient to be determined below through
consistency. In this gauge, after taking Eq.(\ref{f0eqs}) into account,
Eq.(\ref{feq}) reduces to
\begin{equation}
Q_{11}^{0}\phi_{22}-2Q_{12}^{0}\phi_{12}=-\frac{4\gamma+ad}{d-2}\phi_{0}%
\Delta. \label{consistent}%
\end{equation}
Hence, demanding that both the $\delta\bar{\phi}_{0}$ and the $\delta\bar
{\phi}_{11}$ variations be compatible in the gauge $\phi_{11}=\gamma\phi_{0}$
requires that the remaining degrees of freedom $\phi_{12},\phi_{22}$ satisfy
the equation above. In addition, the variation of the original action with
respect to $\phi_{12},\phi_{22}$ demand the equations of motion%
\begin{equation}
Q_{11}^{0}\phi_{0}=0,\;\;Q_{12}^{0}\phi_{0}=0.
\end{equation}
If we work on mass shell for these two equations, we can choose gauges for the
corresponding fields $\phi_{12},\phi_{22},$ and those gauge choices must be
consistent with Eq.(\ref{consistent}). In this way $\phi_{11},\phi_{12}%
,\phi_{22}$ get all determined in terms of $\phi_{0}.$

We now return to $\phi_{0},$ which satisfies the equations $Q_{11}^{0}\phi
_{0}=Q_{12}^{0}\phi_{0}=Q_{22}^{0}\phi_{0}+\phi_{0}\Delta=0.$ We examine how
the dynamics of $\phi_{0}$ can be derived consistently from the gauge fixed
action. After satisfying the $\phi_{12},\phi_{22}$ equations (namely
$Q_{11}^{0}\phi_{0}=Q_{12}^{0}\phi_{0}=0$) as above, and inserting the gauge
$\phi_{11}=\gamma\phi_{0},$ the gauge fixed action in Eq.(\ref{action})
becomes a functional of only $\phi_{0}.$ So the gauge fixed action is%
\begin{equation}
S\left(  \phi_{0}\right)  =\gamma\int\left(  d^{d+2}X\right)  \left[
2\bar{\phi}_{0}Q_{22}^{0}\phi_{0}-\left(  2+\frac{a}{\gamma}\right)
\tilde{\lambda}\left(  \bar{\phi}_{0}\phi_{0}\right)  ^{\frac{d}{d-2}}\right]
. \label{gaugefixed}%
\end{equation}
The variation of this action for $\bar{\delta}\phi_{0}$ gives the equation of
motion%
\begin{equation}
Q_{22}^{0}\phi_{0}+\left(  1+\frac{a}{2\gamma}\right)  \frac{d}{d-2}\phi
_{0}\Delta=0.
\end{equation}
For consistency this must agree with Eq.(\ref{f0eqs}). Therefore, we must
require that the parameter $\gamma$ that appeared as part of the gauge fixing
in Eq.(\ref{gauge11}) must be determined consistently so that $\left(
1+\frac{a}{2\gamma}\right)  \frac{d}{d-2}=1,$ hence%
\begin{equation}
\gamma=-\frac{1}{4}da.
\end{equation}
Thus, the gauge fixed action takes the form%
\begin{equation}
S\left(  \phi_{0}\right)  =-\frac{ad}{2}\int\left(  d^{d+2}X\right)  \left[
\bar{\phi}_{0}Q_{22}^{0}\phi_{0}-\left(  1-\frac{2}{d}\right)  \tilde{\lambda
}\left(  \bar{\phi}_{0}\phi_{0}\right)  ^{\frac{d}{d-2}}\right]  .
\label{actiongauged}%
\end{equation}

In addition to the equation of motion $Q_{22}^{0}\phi_{0}+\phi_{0}\Delta=0$
that follows from this gauge fixed action, we must also impose the conditions
$Q_{11}^{0}\phi_{0}=Q_{12}^{0}\phi_{0}=0.$ So, let us determine the action
after these kinematic equations have been imposed. In section (\ref{cov}) we
have already shown that the solution to $Q_{11}^{0}\phi_{0}=Q_{12}^{0}\phi
_{0}=0$ is
\begin{equation}
\phi_{0}\left(  X\right)  =\delta\left(  X^{2}\right)  \left(  X^{+^{\prime}%
}\right)  ^{-\left(  d-2\right)  /2}F\left(  \frac{X^{-^{\prime}}%
}{X^{+^{\prime}}},\frac{X^{\mu}}{X^{+^{\prime}}}\right)
\end{equation}
and that $Q_{22}^{0}\phi_{0}$ takes the form%
\begin{equation}
Q_{22}^{0}\phi_{0}=-\frac{1}{2}\delta\left(  X^{2}\right)  \frac{\partial^{2}%
}{\partial X^{M}\partial X_{M}}\left[  \left(  X^{+^{\prime}}\right)
^{-\left(  d-2\right)  /2}F\left(  \frac{X^{-^{\prime}}}{X^{+^{\prime}}}%
,\frac{X^{\mu}}{X^{+^{\prime}}}\right)  \right]  .
\end{equation}
After the change of variables $X^{M}\rightarrow\left(  \kappa,\lambda,x^{\mu
}\right)  $ in Eq.(\ref{massless}) this simplifies as in Eq.(\ref{AX})%
\begin{equation}
Q_{22}^{0}\phi_{0}=\delta\left(  \left(  \lambda-\frac{x^{2}}{2}\right)
\kappa^{2}\right)  \kappa^{-d+2}\frac{1}{2\kappa^{2}}\left(  -\frac
{\partial^{2}\varphi\left(  x\right)  }{\partial x^{\mu}\partial x_{\mu}%
}\right)  ,
\end{equation}
where $x^{\mu}$ are the coordinates in $\left(  d-1\right)  +1$ dimensions and
$\left(  \lambda,\kappa\right)  $ are the extra coordinates. In these
coordinates the first term in the Lagrangian takes the form%
\begin{equation}
\bar{\phi}_{0}Q_{22}^{0}\phi_{0}=\sigma\delta\left(  \lambda-\frac{x^{2}}%
{2}\right)  \kappa^{-\left(  d+2\right)  }\left(  -\frac{1}{2}\bar{\varphi
}\left(  x\right)  \frac{\partial^{2}\varphi\left(  x\right)  }{\partial
x^{\mu}\partial x_{\mu}}\right)
\end{equation}
where $\sigma=\delta\left(  0\right)  $ as defined before. The interaction
term has a similar overall factor that is computed as follows%
\begin{align}
\tilde{\lambda}\left(  \bar{\phi}_{0}\phi_{0}\right)  ^{\frac{d}{d-2}}  &
=\tilde{\lambda}\left(  \delta\left(  X^{2}\right)  \kappa^{-\left(
d-2\right)  /2}\right)  ^{^{\frac{2d}{d-2}}}\left(  \bar{\varphi}%
\varphi\right)  ^{\frac{d}{d-2}}\\
&  =\tilde{\lambda}\sigma^{\frac{4}{d-2}+1}\delta\left(  X^{2}\right)
\kappa^{-d}\left(  \bar{\varphi}\varphi\right)  ^{\frac{d}{d-2}}\\
&  =\sigma\delta\left(  \lambda-\frac{x^{2}}{2}\right)  \kappa^{-\left(
d+2\right)  }\lambda\left(  \bar{\varphi}\varphi\right)  ^{\frac{d}{d-2}},
\end{align}
where $\lambda$ is the finite renormalized constant%
\begin{equation}
\lambda=\tilde{\lambda}\sigma^{\frac{4}{d-2}}. \label{renCoupling}%
\end{equation}

Inserting these results into the gauge fixed action in Eq.(\ref{actiongauged}%
), and using
\begin{equation}
\left(  d^{d+2}X\right)  =\kappa^{d+1}d\kappa d\lambda\left(  d^{d}x\right)
\end{equation}
we obtain the Klein-Gordon theory as follows%
\begin{equation}
S\left(  \varphi\right)  =\left[  -\frac{ad}{4}\sigma\left(  \int
d\kappa\right)  \right]  \int d\lambda\delta\left(  \lambda-\frac{x^{2}}%
{2}\right)  \int\left(  d^{d}x\right)  L\left(  \varphi\right)
\end{equation}
where%
\begin{equation}
L\left(  \varphi\right)  =\int\left(  d^{d}x\right)  \left[  -\bar{\varphi
}\frac{\partial^{2}\varphi}{\partial x^{\mu}\partial x_{\mu}}-2\left(
1-\frac{2}{d}\right)  \lambda\left(  \bar{\varphi}\varphi\right)  ^{\frac
{d}{d-2}}\right]  .
\end{equation}
For the correct normalization of the Klein-Gordon field $\varphi\left(
x\right)  $ we must renormalize the overall action by the inverse of the
overall factor $\left[  -\frac{ad}{4}\sigma\left(  \int d\kappa\right)
\right]  $, or else choose the overall extra parameter $a$ that appeared in
$U\left(  \bar{\phi}_{0}\phi_{0}\right)  $ to tune this overall factor to
exactly $1,$ so that%
\begin{equation}
S\left(  \varphi\right)  =\int\left(  d^{d}x\right)  L\left(  \varphi\right)
.
\end{equation}
$L\left(  \varphi\right)  $ is the usual conformally invariant
\textit{interacting} theory with SO$\left(  d,2\right)  $ symmetry at the
classical level. The interacting equations of motion derived from this 1T
action are in agreement with those derived from the original 2T action.

This section developed the methods for gauge fixing the interacting 2T field
theory, and one can now apply similar methods to study other gauges, such as
the ones indicated in Fig.1.

\section{Comments}

We have successfully constructed a field theoretic formulation of 2T-physics
including interactions. The approach is a Sp$\left(  2,R\right)  $ BRST
formulation \cite{ib+yank} based on the underlying Sp$\left(  2,R\right)  $
gauge symmetry of 2T-physics. Some steps were similar to string field theory
\cite{witten} and this was used mainly for inspiration. However, the form of
the interaction and the methods used are new as a BRST gauge theory
formulation. In this we were guided by the previous work at the equation of
motion level \cite{2tfield}. We found that the interaction at the level of the
action is uniquely determined by the interacting BRST gauge symmetry which is
different than the free BRST gauge transformation.

We have shown that with a particular gauge choice the 2T field theory comes
down to the standard interacting Klein-Gordon form. As expected from previous
work in 2T-physics, we have related the global SO$\left(  d,2\right)  $
conformal symmetry of the 1T Klein-Gordon theory directly to the global
space-time symmetry of the 2T-physics field theory in $d+2$ dimensions,
including interactions. Note that the Klein-Gordon theory is just one of the
possible holographic images of the 2T field theory. We should be able to
derive other holographic images, such as those that appear in Fig.1. The
various interacting field theories must then be dual to each other. Through
such dualities we may be able to obtain non-perturbative information about
various field theories.

The BRST operator we started with included only the constraints $Q_{ij}^{0}$
of the worldline action in Eq.(\ref{2Taction}) and include also the
corresponding ghosts $c^{ij},b_{ij}$ (equivalently $c^{m},b_{m}$). It is
possible to enlarge our BRST\ operator by also including the constraints due
to the fact that the gauge fields $A^{ij}$ have vanishing canonical conjugates
$\pi_{ij}=0,$ and the corresponding ghosts $\tilde{c}^{ij},\tilde{b}_{ij}$. If
we had taken that approach\footnote{For a study of this type, but for the much
simpler case of the 1T worldline spinless particle, see \cite{holten}.}, then
the field $\Phi$ would have to be taken as a function of more variables
$\Phi\left(  X,c^{ij},A_{ij},\tilde{c}^{ij}\right)  .$ This would make our
analysis more complete but much more complex, and possibly with no benefit.
However, it may still be useful to pursue this more complete approach.
Similarly, after getting motivated by the transformation properties of all the
fields in the free theory given by $\Phi$ of Eq.(\ref{bigF}), we decided to
concentrate only on $\phi_{0}$ and $\phi_{ij}$ in constructing the interacting
theory, assuming the others decouple. Indeed we found a consistent gauge
theory with the smaller set of fields $\phi_{0}$ and $\phi_{ij}$. It may be
useful to extend our methods to include all the fields in $\Phi$ and then
rewrite the interacting theory in terms of the package $\Phi\left(
X,c^{ij}\right)  .$ Since this does not seem to be urgent we have left it to
future work.

The interacting action that we have suggested provides the first step for
studying 2T-physics at the quantum level through the path integral. This may
be done without reference to 1T-physics, which would be interesting in its own
right, but for interpretation of the results it would be desirable to also
consider gauge fixing to 1T-physics.

The steps that we have performed in the gauge fixing process were done at the
classical field theory level. It is evident that we can study the
corresponding gauge choices in the path integral to relate 2T physics to many
versions of 1T physics, including interactions at the quantum field theory
level. This remains to be investigated in detail, but we expect to be able to
establish physically similar results to the classical version, modulo some
possible quantum corrections in various gauges.

The methods used in this paper can be generalized to spinning particles,
including gauge fields, to find the general interacting 2T field theory. For
this one can use as a guide the previous work at the equation of motion level
for interacting spinning particles up to spin 2 \cite{2tfield}. The
Chern-Simons version of interactions may be relevant in the case of gauge fields.

Our motivation in developing the current formalism was to work our way in a
similar fashion toward building a consistent interacting 2T field theory and
to apply the results to construct the Standard Model of Elementary Particles
and Forces from the point of view of 2T-physics in a BRST field theory
formalism. In fact, the BRST gauged fixed version in Eq.(\ref{actiongauged})
served as a spring board for developing a closely related but simpler
formulation of 2T quantum field theory that applies to spins 0,1/2,1,2. This
provided a short cut for successfully constructing the Standard Model in the
framework of 2T-physics, thus establishing that 2T-physics provides the
correct description of Nature from the point of view of 4+2 dimensions
\cite{2tSM}. The BRST approach may still provide guidance in the long run to
settle remaining issues, as well as to go beyond the Standard Model and toward M-theory.

\section*{Acknowledgments}

We gratefully acknowledge discussions with S-H. Chen, B. Orcal, and G.
Quelin.
%%%%%%%%%%%%%%%%%%%%%%%%%%%%%%%%%%%%%%%%%%%%%%%%%%%%%%%%%%%%%%%%%%%%%%%%%%%%

\end{document}